\documentclass[a4paper,twocolumn,keywords,synopsis]{iucr}
\usepackage[latin9]{inputenc}
\usepackage{textcomp}
\usepackage{amsmath}
\usepackage{amssymb}
\usepackage{graphicx}
\usepackage{esint}
\makeatletter

     \paperprodcode{a000000}      
     \paperref{xx9999}            
     \papertype{FA}               

     \paperlang{english}          
     \journalcode{A}             
     \journalyr{2020}
     \journalreceived{\relax}
     \journalaccepted{\relax}
     \journalonline{\relax}

\makeatother

\begin{document}
\title{Resolution of a bent-crystal spectrometer for X-ray free electron
laser pulses: diamond vs. silicon}
\shorttitle{Resolution of bent-crystal spectrometer}
\author[a]{Vladimir M.}{Kaganer}
\author[b]{Ilia}{Petrov}
\author[b]{Liubov}{Samoylova}
\aff[a]{Paul-Drude-Institut für Festkörperelektronik, Leibniz-Institut im
Forschungsverbund Berlin e.\,V., Hausvogteiplatz 5--7, 10117 Berlin,
Germany}
\aff[b]{European XFEL GmbH, Holzkoppel 4, 22869 Schenefeld, Germany}
\shortauthor{Kaganer, Petrov, and Samoylova}
\maketitle
\begin{synopsis}
Resolution function of a bent-crystal spectrometer for pulses of an
X-ray free electron laser is evaluated. Diffraction on a strongly
bent diamond single crystal gives rise to a resolution of $\Delta E/E=3\times10^{-6}$,
while the strongly bent silicon provides a three times worse resolution. 
\end{synopsis}
\begin{abstract}
The resolution function of a spectrometer based on a strongly bent
single crystal (bending radius of 10~cm or less) is evaluated. It
is shown that the resolution is controlled by two parameters, (i)
the ratio of the lattice spacing of the chosen reflection to the crystal
thickness and (ii) a single parameter comprising crystal thickness,
its bending radius, and anisotropic elastic constants of the chosen
crystal. Diamond, due to its unique elastic properties, can provide notably
higher resolution than silicon. The results allow to optimize the
parameters of bent crystal spectrometers for the hard X-ray free electron
laser sources. 
\end{abstract}

\section{Introduction}

The self-amplified spontaneous emission (SASE) radiation pulse of
an X-ray free-electron laser (XFEL) originates from random current
fluctuations in the electron bunch and have an individual time structure
\cite{saldin:book}. Detectors that could resolve the time structure
of a femtosecond pulse are neither available nor expected in the near
future. The measurement of the energy spectrum of an individual pulse
provides information on its time structure.

A pulse consists of 0.1~fs spikes, which results in an energy range
to be covered by a spectrometer of about 40~eV. A pulse duration
of 50~fs gives rise to a required energy resolution of 0.08~eV.
These requirements are matched by using Bragg diffraction on a thin
single crystal bent to a radius of 10~cm or less. Spectrometers based
on bent silicon \cite{zhu12,makita15} and diamond \cite{boesenberg17,samoylova19}
crystals have been reported so far.

Recently, we have given a detailed theoretical description of the
x-ray diffraction on strongly bent crystals \cite{kaganer20aca}.
It has been proven that the kinematical diffraction approximation
can be applied in a broad range of the crystal curvatures and the
X-ray energies. The diffracted intensity has been calculated. In modeling
diffraction of the XFEL pulses, it has been presumed that the scattering
amplitudes for different frequencies add up coherently.

In the present work, we thoroughly analyze the diffracted intensity
integrated over the pulse duration. We find that the coherent sum
of the amplitudes describes an instant diffraction signal. In the
time-integrated signal, the intensities, rather than the amplitudes,
of different frequencies add up. \citeasnoun{afanasev-kohn77} arrived
at a similar conclusion when analyzing diffraction from a continuous
incoherent X-ray source and averaging over random time instants of
the emission of individual atoms.

We show that the spectral resolution of a bent-crystal spectrometer
is controlled by two parameters. One parameter is simply the ratio
of the lattice spacing of the actual reflection to the crystal thickness.
The other parameter depends on the crystal thickness, its bending
radius, and the anisotropic elastic constants of the crystal. The
elastic properties of diamond make this material particularly suitable
for a bent-crystal spectrometer.

\section{Time-integrated diffraction intensity}

The transverse coherence length of an XFEL pulse is about 1~mm, large
compared to the size of the diffraction region on a strongly bent
crystal. Hence, we assume full transverse coherence of the incident
XFEL pulse and take into consideration only its time structure. The
electric field of the pulse can be represented by its spectrum 
\begin{equation}
E^{\mathrm{in}}(\mathbf{r},t)=\intop_{-\infty}^{\infty}\tilde{E}^{\mathrm{in}}(\omega)e^{ik\mathbf{s}_{0}\cdot\mathbf{r}-i\omega t}\mathrm{\,d}\omega.\label{eq:1a}
\end{equation}
Here $\omega$ is the frequency of a plane-wave component, $k=\omega/c$
is its wavevector, $c$ is the speed of light, and $\mathbf{s}_{0}$
is the unit vector in the direction of the wave propagation.

The wave packet (\ref{eq:1a}) is incident on a bent-crystal spectrometer.
It has been proven by \citeasnoun{kaganer20aca} and discussed below,
that the Bragg diffraction in a strongly bent crystal can be described,
in a wide range of the X-ray energies and bending radii, in the kinematical
(first Born) approximation. In this approximation, the amplitude of
the scattered wave is {[}see, e.g., \citeasnoun{born-wolf19}, Sec.~13.1.2{]}
\begin{equation}
E^{\mathrm{out}}(r\mathbf{s},t)=\frac{r_{e}}{r}\intop_{-\infty}^{\infty}\tilde{E}^{\mathrm{in}}(\omega)e^{ikr-i\omega t}f_{1}(\mathbf{s},\mathbf{s}_{0};k)\mathrm{\,d}\omega.\label{eq:2}
\end{equation}
Here $r_{e}$ is the classical radius of electron and $\mathbf{s}$
is the unit vector in the direction to the detector. The distance
$r$ from the bent-crystal spectrometer to a detector is assumed large
enough, so that the conditions of the Fraunhofer diffraction are satisfied.
The scattering amplitude in the first Born approximation is 
\begin{equation}
f_{1}(\mathbf{s},\mathbf{s}_{0};k)=\int_{V}\varrho(\mathbf{r}')e^{-ik(\mathbf{s}-\mathbf{s}_{0})\cdot\mathbf{r}'}\,\mathrm{d}\mathbf{r}',\label{eq:2a}
\end{equation}
the integration is performed over the crystal volume, and $\varrho(\mathbf{r}')$
is the electron density of the crystal. For X-rays, it is not averaged
over the physically infinitely small volumes but possesses the crystal
lattice periodicity {[}see, e.g., \citeasnoun{landau:electrodyn},
§124{]}. We have restored in Eq.~(\ref{eq:2}) the time exponent
$\exp(-i\omega t)$, which is usually omitted when considering diffraction
of a monochromatic wave, and explicitly noted in Eq.~(\ref{eq:2a})
the dependence of the scattering amplitude $f_{1}$ on the length
of the wavevector $k$.

The intensity of the scattered wave at the time instant $t$ is 
\begin{eqnarray}
I(r\mathbf{s},t) & = & \left|E^{\mathrm{out}}(r\mathbf{s},t)\right|^{2}\nonumber \\
 & = & \frac{r_{e}^{2}}{r^{2}}\iintop_{-\infty}^{\infty}\mathrm{d}\omega_{1}\mathrm{d}\omega_{2}\,e^{i(\omega_{2}-\omega_{1})t}e^{i(k_{1}r-k_{2}r)}\label{eq:2b}\\
 &  & \times\tilde{E}^{\mathrm{in}}(\omega_{1})\tilde{E}^{\mathrm{in}*}(\omega_{2})f_{1}(\mathbf{s},\mathbf{s}_{0};k_{1})f_{1}^{*}(\mathbf{s},\mathbf{s}_{0};k_{2}),\nonumber 
\end{eqnarray}
where the asterisk denotes the complex conjugate and $k_{n}=\omega_{n}/c$
are the wavevectors ($n=1,2$).

Available X-ray detectors cannot resolve the time structure within
the pulse duration (otherwise, the time structure of the pulse would
be measured directly without a spectrometer). Hence, the measured
intensity is a result of integration over the pulse duration: 
\begin{equation}
\mathcal{I}(r\mathbf{s})=\intop_{-\infty}^{\infty}I(r\mathbf{s},t)\,\mathrm{d}t.\label{eq:4}
\end{equation}
Integration of the time-dependent term in Eq.~(\ref{eq:2b}) gives
rise to a delta-function $\delta(\omega_{1}-\omega_{2})$, so that
the intensity integrated over the time is 
\begin{equation}
\mathcal{I}(r\mathbf{s})=\frac{2\pi r_{e}^{2}}{r^{2}}\intop_{-\infty}^{\infty}\left|f_{1}(\mathbf{s},\mathbf{s}_{0};k)\right|^{2}\left|\tilde{E}^{\mathrm{in}}(\omega)\right|^{2}\mathrm{\,d}\omega.\label{eq:5}
\end{equation}
This equation replaces Eq.~(17) by \citeasnoun{kaganer20aca},
where a coherent superposition of the waves with different wave vectors
has been presumed.

Let us calculate now the wave vector transfer $k\mathbf{s}-\left(k\mathbf{s}_{0}+\mathbf{Q}\right)$
for Bragg diffraction at the reciprocal lattice vector $\mathbf{Q}$.
Let the scattering plane be the $xz$ plane with the $x$ axis tangent
to the surface of the bent crystal at $x=0$ and the $z$ axis along
the inner surface normal. All plane wave components of the X-ray pulse
are incident onto the crystal at the same angle $\bar{\theta}_{\mathrm{B}}$
with respect to $x$ axis.

Then, the wavevector of the incident wave is 
\begin{equation}
\mathbf{K}^{\mathrm{in}}=k\mathbf{s}_{0}=k\left(\cos\bar{\theta}_{\mathrm{B}},\sin\bar{\theta}_{\mathrm{B}}\right).\label{eq:5a}
\end{equation}
The angle $\bar{\theta}_{\mathrm{B}}$ is the Bragg angle for a reference
frequency $\bar{\omega}$ arbitrarily chosen in the pulse spectrum.
The Bragg law reads $d\sin\bar{\theta}_{\mathrm{B}}=\pi/\bar{k},$where
$\bar{k}=\bar{\omega}/c$ is the wavevector for the reference frequency
and $d$ is the lattice spacing of the chosen reflection.

The diffracted intensity is measured as a function of the angle $\theta$
between the $x$ axis and the vector $\mathbf{s}$. Hence, the wavevector
of the diffracted wave is 
\begin{equation}
\mathbf{K}^{\mathrm{out}}=k\mathbf{s}=k\left(\cos\theta,-\sin\theta\right).\label{eq:5b}
\end{equation}
With the reciprocal lattice vector $\mathbf{Q}=\left(0,-2\bar{k}\sin\bar{\theta}_{\mathrm{B}}\right)$,
the Bragg law $\mathbf{K}^{\mathrm{out}}=\mathbf{K}^{\mathrm{in}}+\mathbf{Q}$
is satisfied at the reference frequency $\bar{\omega}$. For all other
frequencies presented in the incident pulse, the deviations from the
Bragg law $\mathbf{q}=\mathbf{K}^{\mathrm{out}}-\left(\mathbf{K}^{\mathrm{in}}+\mathbf{Q}\right)$
can be calculated using Eqs.~(\ref{eq:5a}) and (\ref{eq:5b}).

It is convenient to consider the scattering angle $\bar{\theta}_{\mathrm{B}}+\theta$
as twice the Bragg angle of a wave with the frequency $\omega'$ defined
by this condition. The Bragg law reads $d\sin\left[\left(\bar{\theta}_{\mathrm{B}}+\theta\right)/2\right]=\pi/k'$,
where $k'$ is the respective wavevector. A straightforward calculation
{[}see also Appendix B by \citeasnoun{kaganer20aca}{]} gives $q_{z}=2(k'-k)\sin\bar{\theta}_{\mathrm{B}}$.
We will see in the next section that $q_{x}$ is not involved in further
calculations.

Therefore, the squared scattering amplitude $\left|f_{1}(\mathbf{s},\mathbf{s}_{0};k)\right|^{2}$
in Eq.~(\ref{eq:5}) is a function of the difference $\omega'-\omega$.
We denote this function (with the factor $2\pi r_{e}^{2}/r^{2}$ included
in it) as $R(\omega'-\omega)$ and rewrite Eq.~(\ref{eq:5}) as a
convolution integral 
\begin{equation}
\mathcal{J}(\omega')=\intop_{-\infty}^{\infty}R(\omega'-\omega)\left|\tilde{E}^{\mathrm{in}}(\omega)\right|^{2}\mathrm{\,d}\omega.\label{eq:6}
\end{equation}
Here $\mathcal{J}(\omega')$ is the intensity (\ref{eq:5}) after
the change of variables from $\theta$ to $\omega'$. One can see
that if the resolution is ideal {[}i.e., $R(\omega'-\omega)$ is a
delta function{]}, the spectrum of the diffracted waves in the $\omega'$
scale coincides with the spectrum of the incident wave. That justifies
the choice of the variables. Calculation of the function $R(\omega'-\omega)$
for Bragg diffraction from a bent crystal is performed in the next
section.

\section{Resolution of a bent-crystal spectrometer}

The applicability limits of the kinematical approximation to Bragg
diffraction from strongly bent crystals have been established by \citeasnoun{kaganer20aca}.
It has been shown that a bent crystal diffracts kinematically if the
bending radius is so small that the X-ray beam remains at diffraction
condition, i.e., withing the Darwin width of the respective reflection
$\Delta\theta_{\mathrm{B}}$, over a distance in the direction of
its propagation small compared to the extinction length $\Lambda$.
This condition is satisfied as long as the bending radius is small
compared to a critical radius $R_{c}=\Lambda/\Delta\theta_{\mathrm{B}}$.
For hard X-rays with energies larger than 8~keV, the kinematical
approximation is applicable for the bending radii below 10~cm and
the usually employed reflections of silicon and diamond.

The bending gives rise to a displacement field $\mathbf{u}(\mathbf{r})$
in the crystal. For the symmetric Bragg reflections considered in
the present work, only $u_{z}$ component of the displacement field
is of interest. For a crystal cylindrically bent to a radius $R$,
it is $u_{z}=(x^{2}+\alpha z^{2})/2R$. To achieve a cylindrical bending
of a rectangular plate, the bending momenta have to be applied to
the perpendicular edges of the plate. The same bending state can be
approached by applying momentum to the apex of a triangle-shaped plate
\cite{terentjev16}. The parameter $\alpha$ depends on the anisotropic
elastic constants of the crystal \cite{kaganer20aca}. Particularly,
for a 110 oriented diamond plate, $\alpha=0.02$, while for a silicon
plate of the same orientation $\alpha=0.18$. An exceptionally small
value for the diamond, which is a result of compensation of the Poisson
and the anisotropy effects, gives rise to a little depth dependence
of the lattice spacing.

The displacement of atoms due to the bending changes the electron
density of the crystal from a periodic function $\varrho(\mathbf{r})$
for a perfect non-bent crystal to $\varrho\left(\mathbf{r}-\mathbf{u}(\mathbf{r})\right)$.
The Fourier component of the electron density for the actual reflection
$\mathbf{Q}$ changes from $\varrho_{\mathbf{Q}}$ to $\varrho_{\mathbf{Q}}\exp\left[-i\mathbf{Q}\cdot\mathbf{u}(\mathbf{r})\right]$.

The kinematical diffraction amplitude (\ref{eq:2a}) can be written
as an integral over the scattering plane of the crystal 
\begin{equation}
f_{1}=\intop_{-\infty}^{\infty}\mathrm{d}x\intop_{-D/2}^{D/2}\mathrm{d}z\,\exp(iq_{x}x+iq_{z}z-i\mathbf{Q}\cdot\mathbf{u}),\label{eq:1}
\end{equation}
where $D$ is the thickness of the crystal plate. We omit here and
in the next equation the constant prefactors that are not relevant
to our study.

Since the $x$-dependence of the displacement field of a bent crystal
is $x^{2}/2R$, an $x$-range relevant to diffraction is comparable
with $D$ and the integration over $x$ can be extended to the infinite
limits. This integration results in a phase factor which drops out
when calculating $\left|f_{1}\right|^{2}$. In the remaining integral
over $z$, we proceed to a dimensionless variable $\xi=2z/D$. Then,
the integral (\ref{eq:1}) gives 
\begin{equation}
R(\omega'-\omega)=\left|\intop_{-1}^{1}\exp(if\xi-ib\xi^{2})\,\mathrm{d}\xi\right|^{2},\label{eq:7}
\end{equation}
where it is denoted 
\begin{equation}
b=\frac{\pi}{4}\frac{\alpha D^{2}}{Rd},\,\,\,\,\,f=\pi\frac{D}{d}\frac{\omega'-\omega}{\bar{\omega}}.\label{eq:8}
\end{equation}
In deriving dimensionless parameters (\ref{eq:8}), we take into account
that $Q=2\pi/d$.

\begin{figure}
\includegraphics[width=0.8\columnwidth]{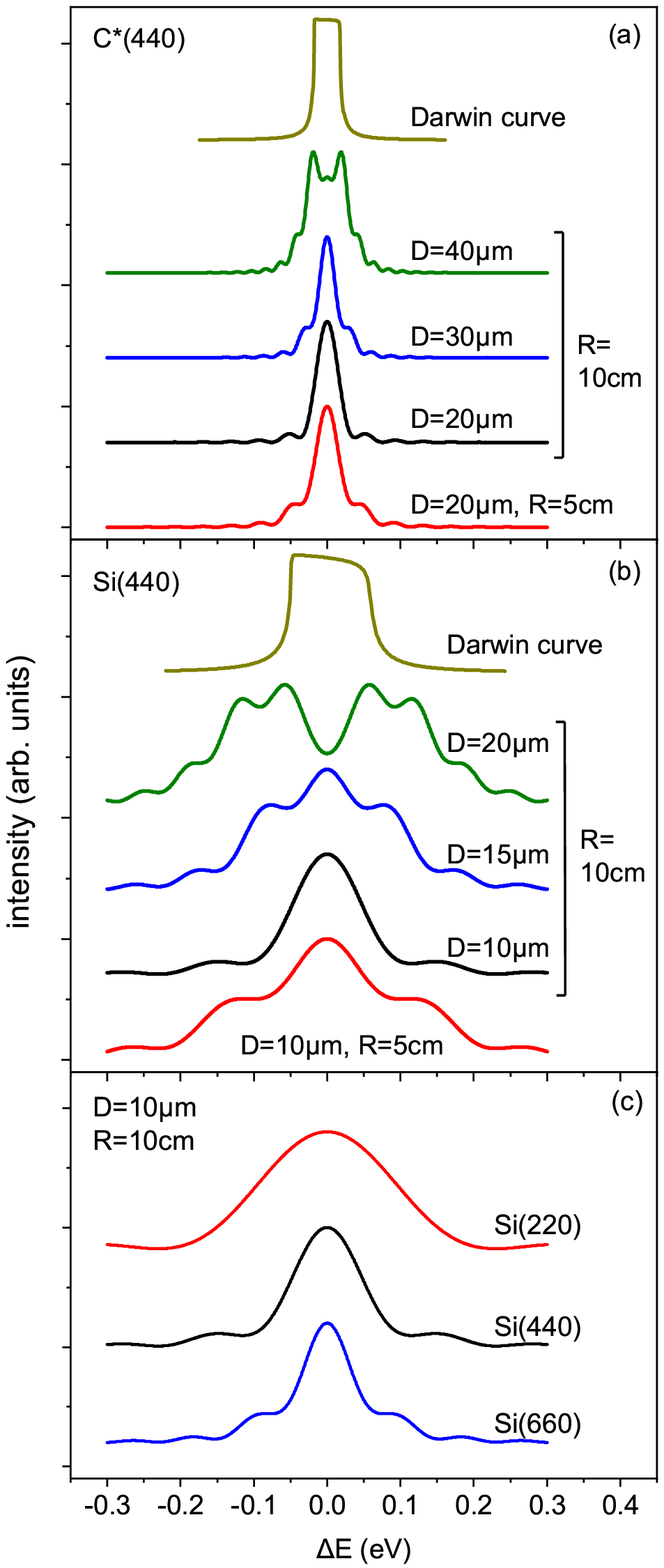}

\caption{Resolution of a bent-crystal spectrometer at the x-ray energy 12~keV
for different bending radii $R$ and crystal thicknesses $D$ for
reflection 440 from diamond (a) and silicon (b), and for different
reflections from silicon (c). Darwin curves of the respective reflections
are shown in (a) and (b).}

\label{fig:resolution} 
\end{figure}

The integral (\ref{eq:7}) can be expressed through cosine and sine
Fresnel integrals $C(x)$ and $S(x)$ as 
\begin{equation}
R(\omega'-\omega)=\left|F\left(\frac{f+b}{\sqrt{2\pi b}}\right)-F\left(\frac{f-b}{\sqrt{2\pi b}}\right)\right|^{2},\label{eq:9}
\end{equation}
where $F(x)=C(x)+iS(x)$. Examples of the resolution function calculated
by Eq.~(\ref{eq:9}) are presented in Fig.~\ref{fig:resolution}.
It is however of interest to investigate the dependence of the resolution
on the parameters $b$ and $f$ qualitatively.

As long as $b\leq1$, the quadratic term in the exponent in Eq.~(\ref{eq:7})
can be neglected, and the resolution function is $\mathrm{sinc}^{2}f$,
where $\mathrm{sinc}(x)=\sin(x)/x$. Then, the resolution by the Rayleigh
criterion is simply $\Delta E/E=(\omega'-\omega)/\bar{\omega}=d/D$,
cf.~Eq.~(16) by \citeasnoun{kaganer20aca}. Let us take a 440
reflection from diamond crystal of thickness $D=20$~\textmu m bent
to a radius of $R=10$~cm and the X-ray energy of $E=12$~keV as
a reference example. In this case, shown in Fig.~\ref{fig:resolution}(a)
by black line, $b=1$. The resolution $\Delta E=0.038$~eV is reached
due to a small value $\alpha=0.02$ for diamond.

The top curve in Fig.~\ref{fig:resolution}(a) is the Darwin curve
of the same reflection, i.e., the rocking curve of dynamical diffraction
curve at an infinitely thick non-bent diamond crystal. It corresponds
to a resolution that could be obtained in a stationary diffraction
experiment by an angular scan of the incoming beam. This resolution
is approximately the same as can be reached by kinematical diffraction
at a strongly bent crystal (black curve), which does not require the
angular scanning and hence is applicable to detect the spectra of
X-ray pulses.

When the bending radius of a 20~\textmu m thick diamond crystal is
reduced to 5~cm or the crystal thickness is increased to 30~\textmu m,
the value of the parameter $b$ increases to $b=2$, and the quadratic
term in the exponent in Eq.~(\ref{eq:7}) cannot be neglected anymore.
The resolution function acquires additional wings, see red and blue
lines in Fig.~\ref{fig:resolution}(a). If the crystal thickness
is increased further to 40~\textmu m, we have $b=4$. Then, the shape
of the resolution function qualitatively transforms, see the olive
curve in Fig.~\ref{fig:resolution}(a). In this case of $b\gg1$,
the integral (\ref{eq:7}) can be calculated by the stationary phase
method. The two peaks in the resolution function are due to two different
depths, above and below the middle plane $z=0$ of the crystal, where
the diffraction takes place depending on the sign of $\Delta E$.
A comparison of the curves in Fig.~\ref{fig:resolution}(a) shows
that one needs to keep $b\leq1$ to reach an optimum resolution. Both
the increase of the curvature and the increase of the crystal thickness
can reduce the resolution.

Figure \ref{fig:resolution}(b) shows the calculated resolution functions
for Si(440) reflection. The crystal thickness $D=10$~\textmu m and
the bending radius $R=10$~cm give $b\approx1.5$ (black curve).
The same range $b\sim1$ is reached at a smaller thickness compared
to the diamond, since $\alpha=0.18$ for silicon is almost an order
of magnitude larger than the respective value $\alpha=0.02$ for diamond.
The resolution is close to the estimate for small\textbf{ $b$}, $\Delta E/E=d/D$,
which gives $\Delta E=0.11$~eV.

The top curve is the Darwin curve for Si(440) reflection at the same
energy of 12~keV. Similarly to the case of diamond above, the resolution
that could be reached by an angular scan in a stationary diffraction
experiment occurs approximately the same as can be reached by the
kinematical diffraction in a strongly bent crystal. The resolution
is three times worse compared to that of the diamond is because of
both a smaller thickness of the bent crystal and a larger lattice
parameter of silicon.

Both a decrease of the curvature radius to $R=5$~cm (red curve)
or an increase of the crystal thickness to $D=15$~\textmu m (blue
curve) causes the increase of the parameter $b$ by a factor of 2
and a decrease of the resolution, roughly by a factor of $\sqrt{2}$.
When the crystal thickness is increased further to $D=20$~\textmu m,
i.e., the thickness that is optimal for the diamond, the resolution
is strongly decreased and the resolution function possesses two peaks
(the olive curve). Thus, the difference in the values of the parameter
$\alpha$, i.e., the difference in anisotropic elastic moduli of the
crystals, results in a smaller silicon crystal thickness needed to
obtain the best available resolution. A smaller thickness, together
with a larger lattice parameter of silicon compared with diamond,
gives rise to a worse achievable resolution.

Figure \ref{fig:resolution}(c) compares the resolution functions
in successive reflection orders of silicon. The parameters providing
an optimum resolution for Si(440), $D=10$~\textmu m and $R=10$~cm,
are used {[}black curves in Figs.\ref{fig:resolution}(a,b){]}. The
Si(220) reflection gives $b=0.73$ and the resolution $\Delta E=0.23$~eV
is worse than that for Si(440) by a factor of 2 because of the two
times larger lattice spacing for this reflection. The Si(660) reflection
provides a narrower curve, but the wings arise because of the increased
value $b=2.2$.

\section{Spectrometer resolution of the XFEL radiation pulses}

\begin{figure}
\includegraphics[width=1\textwidth]{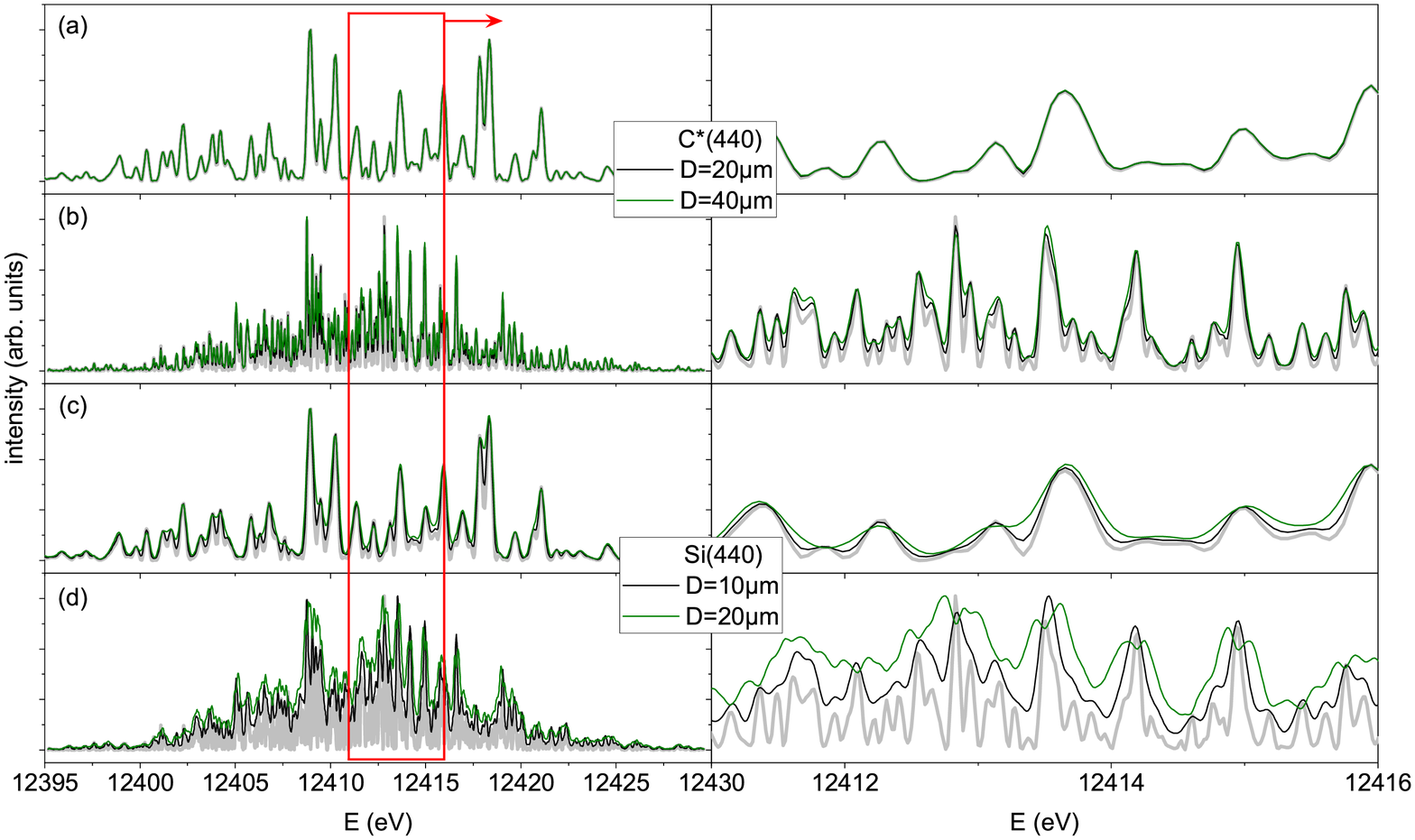}

\caption{Spectra of 10~fs (a,c) and 42~fs pulses (thick gray lines) and the
calculated spectra after diffraction in 440 reflection on (a,b) 20~\textmu m
and 40~\textmu m thick diamond and (c,d) 10~\textmu m and 20~\textmu m
thick silicon crystal plates. The curvature radius is 10~cm.}

\label{fig:spectra} 
\end{figure}

Figure \ref{fig:spectra} compares spectra of the XFEL pulses incident
on the bent crystal with the calculated spectra obtained with the
bent-crystal spectrometers. We employ the same pulses as used in our
former study \cite{kaganer20aca}. The difference is the use of the
convolution integral (\ref{eq:5}) for the time-integrated intensity,
instead of a coherent superposition of the wave with different wave
vectors. The left plots in Figs.~\ref{fig:spectra}(a--d) cover
the whole pulse, while the right ones enlarge a 5~eV wide part of
the spectrum.

The pulses generated during the SASE process at the European XFEL
are simulated with the code FAST \cite{saldin99}, which provides
a 2D distribution of electric field in real space at the exit of the
undulator for each time moment for various parameters of the electron
bunch and the undulator. Simulation results are stored in an in-house
database \cite{manetti19}. The pulses are simulated for the electron
energy 14~GeV, photon energy 12.4~keV, and the active undulator
length corresponding to the saturation length \cite{exfel-fel2014}.
Conversion from the time to the frequency domain is performed using
the WavePropaGator package \cite{samoylova16}, which provides a 2D
distribution of electric field for each frequency of the pulse. We
use the spectrum at the center of the pulse in the frequency domain,
assuming this distribution to be the same across the beam. Two pulses
are compared: a 10~fs pulse generated in an undulator of active length
75~m is used in Figs.~\ref{fig:spectra}(a,c), while a 42~fs pulse
at the undulator length 105~m is used in Figs\@.~\ref{fig:spectra}(b,d).

Each plot in Fig.~\ref{fig:spectra} compares the incident spectrum
(thick gray lines) with two diffracted spectra, corresponding to the
calculated resolution functions presented in Figs.~\ref{fig:resolution}(a,b)
by the same colors. Black curves in Figs.\ref{fig:resolution}(a,b)
and \ref{fig:spectra}(a--d) correspond to the optimal conditions
for the respective reflections: crystal thicknesses $D=20$~\textmu m
for C{*}(440) and $D=10$~\textmu m for Si(440), and the same bending
radius of $R=10$~cm. The olive curves in Figs.\ref{fig:resolution}(a,b)
and \ref{fig:spectra}(a--d) correspond to worse resolutions, obtained
with the crystal thicknesses $D=40$~\textmu m for C{*}(440) and
$D=20$~\textmu m for Si(440), keeping the same bending radius of
$R=10$~cm.

The 10~fs pulse gives rise to peaks in the spectrum with the characteristic
width of 0.35~eV. They are perfectly resolved in the C{*}(440) reflection
in Fig.~\ref{fig:spectra}(a) and reasonably well resolved in the
Si(440) reflection in Fig.~\ref{fig:spectra}(c). Some loss of the
resolution can be seen in the case of Si(440). The 42~fs pulse possesses
0.08~eV wide peaks in the spectrum. The spectrum is well resolved
in the C{*}(440) reflection, see Fig.~\ref{fig:spectra}(b), even
when the crystal thickness is not optimally chosen. The same spectrum
is poorly resolved in Si(440) reflection even with the optimal crystal
thickness and bending radius, see black line in Fig.~\ref{fig:spectra}(d).
The worse resolution at a larger crystal thickness (olive line) notably
affects the diffracted spectrum.

\onecolumn

\twocolumn

\section{Conclusions}

We have shown that the angular distribution of the intensity diffracted
by a bent crystal and integrated over the pulse duration is given
by a convolution of the spectrum of the incident X-ray pulse with
the resolution function of the bent-crystal spectrometer. The resolution
is not affected by the temporal coherence of the pulse.

We have evaluated the resolution of the bent-crystal spectrometer.
It is controlled by two parameters. One parameter is the ratio $d/D$
of lattice spacing of the chosen reflection to the thickness of the
bent crystal. This ratio is the maximum resolution, $\Delta E/E=d/D$,
that is reached if the other parameter, denoted by $b$, is smaller
than 1. The parameter $b$ given by Eq.~(\ref{eq:8}) combines in
a single parameter the crystal thickness $D$, the curvature radius
$R$, the lattice spacing of the actual reflection $d$, as well as
the parameter $\alpha$ representing the anisotropic elastic properties
of the crystal.

For C{*}(440) reflection and a crystal thickness of 20~\textmu m,
the resolution of $\Delta E/E=3\times10^{-6}$ can be reached. For
Si(440), a smaller thickness required to keep the condition $b\leq1$,
together with a larger lattice spacing of silicon, provide a resolution
of $\Delta E/E=9\times10^{-6}$. These results allow to optimize the
parameters of the bent-crystal spectrometers for the XFEL radiation
pulses.

\subsection*{Acknowledgments}

The authors thank Andrei Benediktovitch, Vladimir Bushuev, Leonid
Goray, and Ivan Vartanyants for useful discussions and Timur Flissikowski
for a critical reading of the manuscript.



\end{document}